\documentclass[11pt,a4paper]{article}
\usepackage{amssymb}
\usepackage{amsmath} 
\usepackage{rsfs}    

\newcommand{\bra}{\langle}
\newcommand{\ket}{\rangle}
\newcommand{\vect}[1]{\mbox{\boldmath $ #1 $}}
\newcommand{\R}{{\mathbb R}}

\newcommand{\1}{\mbox{\boldmath $ 1 $}}
\newcommand{\A}{\mathscr{A}}

\newcommand{\Hilbert}{\mathscr{H}}

\newcommand{\G}{\mathscr{G}}
\newcommand{\M}{\mathscr{M}}

\newcommand{\Aut}{\mbox{Aut}}

\newcommand{\mapright}[2]
{\mathop{\hbox to 8mm{\rightarrowfill}}
\limits^{\scriptstyle #1}_{\scriptstyle #2}}

\newcommand{\mapdown}[2]
{\Big \downarrow 
\llap {$\vcenter {\hbox{$\scriptstyle #1 \,$}}$ }
\rlap {$\vcenter {\hbox{$\scriptstyle #2   $}}$ }
}

\makeatletter
\@addtoreset{equation}{section}
\makeatother

\begin{document} 
\baselineskip 6.0mm 
\begin{flushright}
2011 December 24
\\
arXiv:1112.xxxx 
\end{flushright}
\vspace*{8mm}
\begin{center}
{\bf \Large
Superselection Rules from Measurement Theory}
\vspace{6mm}

Shogo Tanimura\footnote{{\tt E-mail: tanimura[AT]is.nagoya-u.ac.jp}}
\vspace{6mm}

{\it
Department of Complex Systems Science,
Graduate School of Information Science
\\
Nagoya University,
Nagoya 464-8601, Japan
}
\vspace{8mm}

Abstract
\vspace{4mm}

\begin{minipage}{120mm}
\baselineskip 5.5mm 
In quantum theory, 
physically measurable quantities of a microscopic system are represented
by self-adjoint operators.
However, not all of the self-adjoint operators correspond to measurable quantities.
The superselection rule is a criterion to distinguish measurable quantities.
Any measurable quantity must obey the superselection rules.
By contraposition,
any quantity which does not obey the superselection rules cannot be measured.
Although some of superselection rules were proved,
the raizon d'{\^e}tre of the superselection rules has been still obscure.
In this paper we deduce the superselection rules 
from an assumption on symmetry property of measurement process.
We introduce the notion of covariant indicator,
which is a macroscopic observable 
whose value indicates the value of a microscopic object observable.
We prove that
if an object system has a quantity that is conserved during the measurement process,
other quantities that do not commute with the conserved quantity
are non-measurable by the covariant indicator.
Our derivation of superselection rules is compared 
with the uncertainty relation under the restriction by a conservation law.
An implication of the color superselection rule for the color confinement is discussed.
It is also argued that spontaneous symmetry breaking enables a measurement 
that the superselection rule prohibits.
\end{minipage}
\end{center}
\vspace*{10mm}
\baselineskip 5.5mm 
\noindent
Subject classes: quant-ph, hep-th
\\

\newpage
\baselineskip 5.7mm 
\section{Introduction}
In the standard formulation of quantum mechanics,
a state of a physical system is represented by a unit ray in a Hilbert space
and 
an observable quantity is represented by a self-adjoint operator acting on the Hilbert space.
However, it is known that
not all of self-adjoint operators in realistic models
correspond to measurable physical quantities.
Although the notion of measurability will be precisely defined later in this paper,
here measurability means ability for making correlation 
between a microscopic quantity to be measured 
and a macroscopic quantity to be read out directly.
If a variation of the value of the microscopic quantity causes
a change of the value of the macroscopic quantity 
via interaction between the microscopic and the macroscopic systems,
we say that we can measure the microscopic quantity.
For example, Millikan determined electric charges of electrons
by measuring velocities of charged oil droplets 
suspended between two metal electrodes in the gravitational field.
In this case, the electron charge causes
a change of motion of the oil droplet,
and he inferred the electron charge
from the data on the motion of the oil droplet.

In the context of quantum field theory,
the electric current $ J^\mu = \bar{\psi} \gamma^\mu \psi $
is defined in terms of 
the Dirac spinor field operator $ \psi $ for electrons.
The electric current $ J^\mu $ is self-adjoint and measurable.
However, the operators
\begin{equation}
	\frac{1}{2} ( \psi + \psi^\dagger ),
	\qquad
	\frac{1}{2i} ( \psi - \psi^\dagger )
\end{equation}
are self-adjoint but
they are not measurable even via indirect methods.

Around 1950 it was puzzling physicists that
the intrinsic parity transformation of spinor field was not uniquely defined.
A phase factor can be multiplied on the parity-transformed spinor field
and it is not uniquely fixed.
Wick, Wightman, and Wigner~\cite{WWW1952} noticed that
the ambiguity in the definition of the parity transformation is allowed since
{\it the spinor field itself is not measurable}.
Thus any choice among possible phase factors does not make changes
in predictions that can be tested by experiments.

{}From their argument physicists have learned that
not every self-adjoint operator
appearing in the formulation of quantum theory
corresponds to a physically measurable quantity.
Hence, we would like to have a criterion
with which we can select measurable operators among all the self-adjoint operators.
Superselection rules work as such criteria.

A superselection rule is stated as follows.
There is an operator $ J $, which is called the superselection charge.
If a self-adjoint operator $ A $ represents a measurable quantity,
it must satisfy the commutativity
\begin{equation}
	[ J, A ] = 0.
	\label{superselection rule}
\end{equation}
This is a superselection rule, 
which is a necessary condition for measurability of $ A $.

The superselection rule can be compared with a conservation law.
The conservation of $ J $ is formulated as
\begin{equation}
	[ J, H ] = 0,
	\label{conservation law}
\end{equation}
where $ H $ is the Hamiltonian $ H $ of the system.
The conservation law (\ref{conservation law}) requires
that $ J $ commutes with the Hamiltonian $ H $
while
the superselection rule (\ref{superselection rule}) requires
that $ J $ commutes with all of the measurable quantities.
Thus the superselection rule is a stronger requirement for $ J $
than the conservation law.
It can be said that 
the superselection rule is an extreme form of conservation laws.

The history of studies of superselection rules
has been reviewed by Wightman~\cite{Wightman1995} in detail.
Here we briefly review the development of studies of superselection rules
not necessarily in chronological order.
Wick, Wightman, and Wigner~\cite{WWW1952} noticed that 
the fields of half-integer spins are non-measurable
and they formulated the univalence superselection rule, 
which forbids measurements of half-integer spin fields.
In their formulation the superselection charge is $ J = (-1)^{2j} $,
where $ j $ is the total angular momentum of the system.
First they proved the univalence superselection rule 
using the time reversal symmetry.
Later Hegerfeldt, Kraus, and Wigner~\cite{Hegerfeldt1968} proved it
using the rotational symmetry
and they identified the superselection charge as 
$ J = R( 2 \pi ) $, a rotation by a $ 2 \pi $ angle around any axis.
Wick, Wightman, and Wigner~\cite{WWW1952} suggested that
the electric charge could be another superselection charge.
Later Strocchi and Wightman~\cite{Strocchi1974} proved it
in the context of quantum electrodynamics.
In general, the Hilbert space of a system is decomposed into
subspaces which belong to distinct eigenvalues of a superselection charge
and each subspace is called a sector.
Ojima~\cite{Ojima2005} refined the notion of sector 
as a quasiequivalence class of factor states of the algebra of measurable quantities.
No measurable quantity has nonvanishing matrix elements
between arbitrary two state vectors that belong to different sectors.
Hence the relative phase of two vectors in different sectors is un-observable 
and the superposition of the two vectors looks like a mixed state 
for any measurements.
This kind of apparent loss of coherence of superposed state vectors
is called decoherence by Zurek~\cite{Zurek1982, Zurek1991, Zurek2003}.
He noted that superselection rules can provide a mechanism which generates
classical behavior from quantum physics. 
Machida and Namiki~\cite{Machida1980} discussed 
the mechanism of reduction of wave packet
and Araki~\cite{Araki1980} showed that
their theory can be formulated in terms of superselection charges
that have continuous spectra.
Doplicher, Haag, and Roberts~\cite{DHR1969} showed
that an abelian group of superselection charges
can be reconstructed from the algebra of observable quantities. 
This argument has been extended to include non-abelian groups~\cite{DR1972, DR1990}.
They also showed that the fermionic fields 
can be reconstructed from the algebra of observable bosonic quantities~\cite{DHR1971}.
On the other hand,
the spectra of superselection charges are not changed 
by operations of microscopic local observables
and hence they play roles of 
macroscopic classical variables or order parameters, which label distinct sectors.
This aspect of superselection charges has been noted by
Wightman~\cite{Wightman1995}, 
Sewell~\cite{Sewell2002},
and Ojima~\cite{Ojima2003}.
Even though Hepp~\cite{Hepp1972} did not use the word ``superselection rule,''
he showed that disjoint representations of a local observable algebra
are parameterized by expectation values of macroscopic observables,
which are equivalent to the superselection charges.
Ojima~\cite{Ojima2005} proposed the Micro-Macro duality, which signifies
bi-directional functions 
between the category of microscopic quantum systems and
the category of macroscopic classical systems.
In the view of the Micro-Macro duality,
a macroscopic system emerges from a microscopic system
while 
the macroscopic system works as a describer,
an interpreter and a controller acting on the microscopic system.
In this picture,
the superselection sector plays a role of border between
microscopic physics and macroscopic physics.

The history of measurement theory is too huge for reviewing.
Here we only mention Ozawa's theory~\cite{Ozawa2004}, 
which axiomatically characterizes physically feasible measurement processes 
and proves that all feasible measurements 
can be described by models of the von Neumann type.
Hence, we can use von Neumann models without excluding other possibilities.
On the other hand,
in the theory of measurement,
it had been a subtle problem to define 
equality of two observables which belong to distinct subsystems
or
equality of two observables which are defined at different times.
Ozawa~\cite{Ozawa2006} proposed 
saying that two observables have perfect correlation 
if the joint probability distribution 
of outcomes of simultaneous measurements of the two observables is well defined 
and moreover
if the probability for obtaining different outcomes of the two observable is zero.
In this paper we call the perfect correlation Ozawa equality.

Let us turn our attention to the subject of this paper.
Superselection rules often take forms of forbidding rules.
For example,
a quantity that is variant under $ 2 \pi $ rotation must not be measured,
and
a gauge variant quantity must not be measured.
However, the reason why those measurements are impossible is still vague.
The purpose of this paper is to explain a mechanism
that makes those measurements impossible.
We deduce a general superselection rule
as a consequence of symmetries of measurement processes.
More concretely,
we derive
the superselection rule (\ref{superselection rule}) for arbitrary measurable quantities
from the conservation law (\ref{conservation law}) for the superselection charge.

In this paper,
we will begin our discussion by examining simple examples
and show that 
the conservation law of the momentum of an object system prevents
the position measurement.
For formulating the general problem, we will introduce three notions;
isolated conservation law,
covariant indicator,
and Ozawa equality.
Using these notions, we will prove the main theorem;
only a quantity that commutes with the isolated conserved quantity 
is measurable by a covariant indicator. 
This is the most general form of the superselection rules.
We will compare this result with the Wigner-Araki-Yanase-Ozawa theorem.
We will also discuss implications of superselection rules
for both abelian and non-abelian gauge symmetries.
This discussion may give an insight for understanding of color confinement.

\section{Preliminary studies}
\subsection{Von Neumann model of position measurement}
To see an example in which
a measurement and a conservation law are incompatible,
let us investigate
the von Neumann model of position measurement~\cite{Ozawa2004, Neumann1955}.
Assume that we have two systems; 
one is a microscopic object to be observed
and the other is a measuring apparatus.
The object system has a pair of canonical variables $ (q,p) $
and the apparatus has 
another pair of canonical variables $ (Q,P) $.
Suppose that we aim to measure the position $ q $ of the object
by reading the position $ Q $ of the indicator.
In the von Neumann model,
time evolution of the composite system is described by the Hamiltonian 
\begin{equation}
	H_N := K q P
	\label{von Neumann model}
\end{equation}
with a coupling constant $ K $.
The von Neumann Hamiltonian $ H_N $ 
does not have kinetic terms of the respective subsystems
but has only their interaction term.
We take the Heisenberg picture in which the operators change via time evolution
while the state vectors remain unchanged.
Accordingly, the indicator moves as
\begin{equation}
	Q \mapsto 
	\alpha (Q) 
	:= e^{iH_N t / \hbar} \, Q \, e^{-iH_N t / \hbar}
	= Q + q
\end{equation}
when $ K t / \hbar = 1 $.
Here used the canonical commutation relation $ [ Q,P ] = i \hbar $.
Hence, if we know the initial position distribution of $ Q $,
we can infer the object position $ q $
by reading the indicator position $ \alpha (Q) $ 
after the measurement interaction.
It is to be noted that
a mechanism correlating the indicator position to the object position
is necessary for accomplishing a meaningful measurement.

On the other hand, the momentum of the object system changes to
\begin{equation}
	p \mapsto 
	\alpha (p) 
	:= e^{iH_N t / \hbar} \, p \, e^{-iH_N t / \hbar}
	= p - P
\end{equation}
after the measurement interaction.
Here used $ [ q,p ] = i \hbar $.
Even if we replace $ H_N $ by another Hamiltonian $ H $ 
to define a more general model,
the Hamiltonian $ H $ must contain the operator $ q $ to make correlation
between $ q $ and $ \alpha (Q) $.
Hence, $ H $ does not commute with $ p $
and causes a change of the momentum
as $ e^{iHt / \hbar} \, p \, e^{-iHt / \hbar} \ne p $.
Change of the object momentum is unavoidable in any position measurement.
As a contraposition,
we can say that
{\it we cannot measure the object position $ q $
with conserving the object momentum $ p $.}
The position measurement and the momentum conservation 
are incompatible.
Here is a hint for understanding the general superselection rules.

\subsection{Conservation vs. measurement}
Let us investigate another model which exemplifies incompatibility
between conservation and measurement of noncommutative observables.
Assume that 
the object system consists of $ n $ massive particles in the one-dimensional space.
The mass, position, and momentum of each particle are denoted as
$ m_r, q_r, p_r $ $ (r=1, \cdots, n ) $.
The center of mass and the total momentum of the object system are defined as
\begin{equation}
	x := \frac{ \; \sum_{r=1}^n m_r q_r \;}{ \sum_{r=1}^n m_r },
	\qquad
	p_x := \sum_{r=1}^n p_r,
\end{equation}
respectively.
The measuring apparatus has an observable $ M $,
which is called a meter, a pointer, or an indicator.
We would like to design a measurement process
that causes a shift of the meter as
\begin{equation}
	M \mapsto 
	\alpha (M) 
	:= e^{iHt / \hbar} \, M \, e^{-iHt / \hbar}
	= M + x
	\label{meter shift}
\end{equation}
at a specific time $ t $.
Moreover, it is required that the total momentum of the object system is conserved as
\begin{equation}
	p_x \mapsto
	\alpha( p_x ) 
	:= e^{iHt / \hbar} \, p_x \, e^{-iHt / \hbar}
	= p_x.
	\label{momentum conservation}
\end{equation}
Then, it is easily proved that
there is no measurement process $ \alpha $ 
satisfying both the shift property (\ref{meter shift})
and the conservation law (\ref{momentum conservation}).
Since the momentum $ p_x $ is a physical quantity attributed to the object system
and the meter observable $ M $ is attributed to the apparatus,
they commutes, $ [p_x, M] = 0 $. 
The mapping $ \alpha $ describing the time evolution of any physical quantity
$ A \mapsto \alpha (A) = e^{iHt / \hbar} \, A \, e^{-iHt / \hbar}$
is an automorphism of the algebra of observables.\footnote{
In the algebraic formalism of quantum theory,
the algebra of observables can contain
both self-adjoint operators and non-self-adjoint operators.
Since a product of two self-adjoint operators is usually non-self-adjoint,
it is convenient to accept non-self-adjoint elements into the algebra.
A genuine `observable' quantity is demanded to be self-adjoint.}
Hence,
\begin{equation}
	[ \alpha(p_x), \alpha(M) ]
	= \alpha( [p_x, M] ) 
	= 0.
	\label{contradiction 1}
\end{equation}
On the other hand, the center of mass $ x $ and the total momentum $ p_x $ satisfy
the canonical commutation relation
$ [ x, p_x ] = i \hbar $.
Therefore, the two assumptions (\ref{meter shift}) and (\ref{momentum conservation}) 
imply that
\begin{equation}
	[ \alpha(p_x), \alpha(M) ]
	= [ p_x, M + x ]
	= - i \hbar,
	\label{contradiction 2}
\end{equation}
which contradicts (\ref{contradiction 1}).
Hence, there is no Hamiltonian satisfying the two requirements
(\ref{meter shift}) and (\ref{momentum conservation}).

Although 
the requirement (\ref{meter shift}) may be replaced by a more relaxed requirement,
the consequence remains unchanged.
We may use another Hamiltonian $ H $ to define the time evolution
$ \alpha (M) = e^{iHt / \hbar} \, M \, e^{-iHt / \hbar} $.
Instead of (\ref{meter shift}),
we require that the meter observable $ M $ changes to
$ \alpha(M) = f (M,x) $, a nontrivial function of $ x $,
via the measurement process.
In this case, again we have
$ [ \alpha(p_x), \alpha(M) ] = \alpha( [p_x, M] ) = 0 $.
On the other hand, we have
$ [ \alpha(p_x), \alpha(M) ] = [ p_x, f(M,x) ] \ne 0 $
for any nontrivial function $ f(x) $.
Thus it is impossible to design a measurement process $ \alpha $
satisfying both 
the momentum conservation $ \alpha(p_x) = p_x $
and the meter shift condition $ \alpha(M) = f(M, x) $.
We conclude that
{\it any process cannot make a correlation 
between the center-of-mass position of the object system
and the meter position of the apparatus
without violating conservation of the total momentum of the object system.}

The above argument can be generalized for any quantity $ A $ of the object system.
We would like to have 
a measurement process that causes 
the shift of the meter as 
$ M \mapsto \alpha(M) = M + A $.
The momentum conservation implies 
$ [ p_x, H ] = 0 $
and
$ \alpha( p_x ) 
= e^{iHt / \hbar} \, p_x \, e^{-iHt / \hbar}
= p_x $.
Since observables belonging to different subsystems commute,
we have $ [ p_x, M ] = 0 $.
These yield the relation
\begin{equation}
	0 
	= \alpha( [ p_x, M ] )
	= [ \alpha( p_x ), \alpha( M ) ] 
	= [ p_x, M + A ]
	= [ p_x, A ].
\end{equation}
Hence we reach the consequence that any measurable quantity $ A $ must satisfy
\begin{equation}
	[ p_x, A ] = 0.
	\label{momentum selection rule}
\end{equation}
Thus we deduced a superselection rule from the momentum conservation law.
Any quantity that does not commute with $ p_x $
cannot be measured via a momentum-conserving process.
By generalizing this argument, we can derive the superselection rule
$ [ J, A ] = 0 $ of (\ref{superselection rule})
from the conservation law
$ [ J, H ] = 0 $ of (\ref{conservation law}).
This generalization will be established as the main theorem of this paper.

{}For the momentum superselection rule (\ref{momentum selection rule}),
the relative coordinate $ q_s - q_r $ of two particles
commutes with the total momentum $ p_x $.
Hence, $ q_s - q_r $ is measurable.
In a three-particle system,
\begin{equation}
	A = \frac{m_1 q_1 + m_2 q_2}{m_1 + m_2} - q_3
\end{equation}
commutes with
both $ p_1 + p_2 + p_3 $ and $ m_2 p_1 - m_1 p_2 $.
Hence the quantity $ A $ is also measurable
via a momentum-conserving process.

In the above argument,
it is required that the total momentum of the object particles alone is conserved.
On the other hand, in a usual argument
the momentum conservation means that 
the sum of the momenta of the object particles and 
the momentum of another system interacting with the particles
is constant in time duration.
In this paper we propose to call
the conservation of a quantity attributed to only the object system 
an isolated conservation law.

Conservation laws are related to symmetry properties of physical systems.
The analysis presented above suggests that
there is a relation between the symmetry of the measurement process
and the superselection rule.
We will further investigate this point in the following.

\section{Formulation of the problem and the main theorem}
\subsection{Definitions of basic notions}
Here we introduce three notions:
isolated conservation law, 
covariant indicator, and Ozawa equality
to describe the problem under consideration.
We will deduce the superselection rules using these notions.

\subsubsection{Isolated conservation law}
We use the concept of group action
for characterizing correlation between distinct systems.
Suppose that we have two systems, an object system and a measuring apparatus.
The object system has an algebra $ \A $ of its physical quantities
and the apparatus has an algebra $ \M $ of its physical quantities.
The automorphism group of the algebra $ \A $ is denoted as $ \Aut ( \A ) $.
Similarly, 
the automorphism group of the algebra $ \M $ is denoted as $ \Aut ( \M ) $.
Let a group $ G $ act on the two systems.
For making our argument mathematically rigorous, it is safer to assume that
the group $ G $ is a compact Lie group.
The group action on each system is described by
a group homomorphism $ \sigma : G \to \Aut ( \A ), $ $ g \mapsto \sigma_g $
and 
a group homomorphism $ \tau : G \to \Aut ( \M ), $ $ g \mapsto \tau_g $.
We can construct tensor products
$ \sigma_g \otimes \tau_g $,
$ \sigma_g \otimes \mbox{id} $,
$ \mbox{id} \otimes \tau_g $;
all of them are automorphisms of the tensor product algebra
$ \A \otimes \M $ of the composite system.
A measurement process is described by the time-evolution automorphism
$ \alpha : \A \otimes \M \to \A \otimes \M $.
If the diagram
\begin{equation}
	\begin{array}{ccc}
	\A \otimes \M & \mapright{\alpha}{} & \A \otimes \M \\
	\mapdown{\sigma_g \otimes \tau_g \,}{} 
	& & \mapdown{}{\sigma_g \otimes \tau_g} \\
	\A \otimes \M & \mapright{\alpha}{} & \A \otimes \M
	\end{array}
	\label{total conservation law}
\end{equation}
is commutative for arbitrary $ g \in G $,
we say that the measurement process $ \alpha $ is $ G $-invariant
or that the measurement process admits 
the {\it total conservation law}
associated to the action $ \sigma \otimes \tau $ of the group $ G $.
The commutativity of the diagram (\ref{total conservation law})
means that the equation
\begin{equation}
	\alpha (( \sigma_g \otimes \tau_g ) (B))
	=
	( \sigma_g \otimes \tau_g ) ( \alpha (B))
	\label{algebraic commutativity}
\end{equation}
holds for an arbitrary physical quantity $ B \in \A \otimes \M $
and for an arbitrary group element $ g \in G $.

The above definition 
of the $ G $-invariance of the measurement process $ \alpha $
relies on the algebraic formalism of quantum mechanics~\cite{Segal1947b}.
We can reformulate it in the familiar operator formalism as shown below.
Suppose that 
the group $ G $ has generators $ J \in \A $ and $ K \in \M $.
Then the group actions are implemented by unitary transformations as
\begin{equation}
	\sigma_s (A) = e^{iJs / \hbar} A \, e^{-iJs / \hbar},
	\qquad
	\tau_s (M) = e^{iKs / \hbar} M \, e^{-iKs / \hbar}
	\label{unitary implement}
\end{equation}
for arbitrary $ s \in \R $, $ A \in \A $, $ M \in \M $.
On the other hand, the time evolution is described by the Heisenberg operator
\begin{equation}
	\alpha_t (B) = e^{iHt / \hbar} \, B \, e^{-iHt / \hbar},
\end{equation}
where $ H $ is the Hamiltonian of the composite system
and $ B \in \A \otimes \M $ is an arbitrary physical quantity of the composite system.
If the sum $ J + K $ commutes with $ H $ as
\begin{equation}
	[ (J+K), H] = 0,
	\label{conservation of sum}
\end{equation}
then $ J+K $ satisfies the conservation law $ \alpha_t (J+K) = J+K $.
Therefore, the equality
\begin{eqnarray}
	\alpha_t (( \sigma_s \otimes \tau_s ) (B))
	&=& e^{iHt / \hbar} \, e^{i(J+K)s / \hbar} 
	\, B \, e^{-i(J+K)s / \hbar} \, e^{-iHt / \hbar}
\nonumber \\
	&=& e^{i(J+K)s / \hbar} \, e^{iHt / \hbar} 
	\, B \, e^{-iHt / \hbar} \, e^{-i(J+K)s / \hbar}
\nonumber \\
	&=& ( \sigma_s \otimes \tau_s ) ( \alpha_t (B))
	\label{total conserving process}
\end{eqnarray}
holds for an arbitrary $ B \in \A \otimes \M $ and for arbitrary $ s, t \in \R $.
Thus the commutativity (\ref{algebraic commutativity}) is ensured.

If, instead of (\ref{total conservation law}), the diagram
\begin{equation}
	\begin{array}{ccc}
	\A \otimes \M & \mapright{\alpha}{} & \A \otimes \M \\
	\mapdown{\sigma_g \otimes \rm{id} \,}{} 
	& & \mapdown{}{\sigma_g \otimes \rm{id}} \\
	\A \otimes \M & \mapright{\alpha}{} & \A \otimes \M
	\end{array}
	\label{isolated conservation law}
\end{equation}
is commutative for arbitrary $ g \in G $,
we say that the measurement process $ \alpha $
admits the {\it isolated conservation law} associated to
the action $ \sigma $ of the group $ G $.
If the conservation law of $ J $, that is
\begin{equation}
	[J, H] = 0,
	\label{isolated conservation relation}
\end{equation}
holds, the commutativity
$ \alpha_t (( \sigma_s \otimes {\rm id} ) (B))
= ( \sigma_s \otimes {\rm id} ) ( \alpha_t (B)) $
is verified by a calculation similar to (\ref{total conserving process}).
It is to be noted that the quantity $ J $
is attributed to the object system only.
In this case
the conserved quantity $ J $ becomes the superselection charge
as seen below.

Although we introduced the operators $ H, J, K $ 
for making the formulation familiar to physicists
and for showing the conserved quantities explicitly,
we will not use them 
but later we will use only the algebraic relation (\ref{isolated conservation law})
for deriving the superselection rules.

\subsubsection{Covariant indicator}
We have a composite system of the object and the apparatus.
Any quantity $ B $ changes as
$ B \mapsto \alpha (B) = e^{iHt/\hbar} B e^{-iHt/\hbar} $
via a measurement process.
In general,
we read out the meter observable $ \alpha(M) $ of the apparatus
after the measurement process
and infer the value of the object observable $ A $.
To perform a meaningful measurement
we need to make correlation between
the initial object quantity $ A $ and
the indicator $ \alpha(M) $.
In other words,
a change of $ A $ should be followed by a change of $ \alpha(M) $.
Hence, it is appropriate to characterize their correlation 
by the covariance of $ A $ and $ \alpha(M) $ under group transformations.

Let us examine how the covariance is formulated
in the von Neumann model of position measurement.
In that model a shift of the object position by a length $ b \in \R $ 
is described as
\begin{equation}
	\sigma_b (q) 
	:= e^{ipb/\hbar} \, q \, e^{-ipb/\hbar} 
	= q + b.
\end{equation}
A shift of the indicator position is similarly described as
\begin{equation}
	\tau_b (Q) 
	:= e^{iPb/\hbar} \, Q \, e^{-iPb/\hbar} 
	= Q + b.
\end{equation}
In this situation, the covariance of the object and the indicator is characterized
by the condition
\begin{equation}
	\sigma_b ( \alpha (Q) ) =
	\alpha ( \tau_b (Q) ),
	\label{von Neumann covariance}
\end{equation}
which is verified as
\begin{eqnarray}
	\sigma_b ( \alpha (Q) ) 
&=&	e^{ipb/\hbar} \, e^{iH_N t/\hbar} \, Q \, e^{-iH_N t/\hbar} \, e^{-ipb/\hbar} 
	\nonumber \\
&=&	e^{ipb/\hbar} ( Q + q ) \, e^{-ipb/\hbar} 
	\nonumber \\
&=&	Q + (q+b)
	\nonumber \\
&=&	(Q+q) + b
	\nonumber \\
&=&	e^{iH_N t/\hbar} ( Q + b ) \, e^{-iH_N t/\hbar} 
	\nonumber \\
&=&	e^{iH_N t/\hbar} \, e^{iPb/\hbar} \, Q \, e^{-iPb/\hbar} \, e^{-iH_N t/\hbar} 
	\nonumber \\
&=&	\alpha ( \tau_b (Q) )
\end{eqnarray}
for the von Neumann Hamiltonian (\ref{von Neumann model}).
Thus, the shift of the indicator follows 
the shift of the initial position of the object.

By generalizing the above consideration,
we define the notion of a meter observable 
moving covariantly to the object.
If the meter observable $ M $ of the apparatus satisfies
the commutative diagram
\begin{equation}
	\begin{array}{ccc}
	\1 \otimes M & \mapright{\alpha}{} & \alpha( \1 \otimes M ) \\
	\mapdown{\rm{id} \otimes \tau_g \,}{} 
	& & \mapdown{}{\sigma_g \otimes \rm{id}} \\
	\1 \otimes \tau_g M & \mapright{\alpha}{} & 
	\alpha ( \1 \otimes \tau_g M ) = 
	\sigma_g ( \alpha( \1 \otimes M ) )
	\end{array}
	\label{covariant indicator}
\end{equation}
for arbitrary $ g \in G $
and for the identity $ \1 \in \A $,
then $ M $ is called a $ G $-{\it covariant indicator}.
This is a generalization of the shift-covariance property
(\ref{von Neumann covariance}) of the meter.

The notion of covariance of observables under group actions
has been introduced by Holevo~\cite{Holevo}.
However, our definition of covariance is different from his.
In Holevo's definition, 
the covariance means a group transformation property 
of a probability operator-valued measure (POVM)
of a single object system.
In our definition, the covariance means the correlation of 
group transformation properties of two systems.

\subsubsection{Ozawa equality}
In measurement theory,
it had been a subtle issue 
to define equality of two observables belonging to distinct subsystems.
We need to compare $ A $ and $ \alpha(M) $;
$ A $ is an object observable before the measurement process while
$ \alpha(M) $ is a meter observable after the process.
Naively, it seems necessary to make the operator identity
$ A = \alpha(M) $ for carrying out a precise measurement.
However, requiring them to be equal without depending on the state of the system
is an excessive demand.
Once the initial state of the composite system is prepared,
a some part of the spectrum of an observable
is realized as measurement outcomes,
but not all of the spectral values are realized as outcomes with nonzero probability.
Even if some parts of the spectra of $ A $ and $ \alpha(M) $ are different,
if their realization probabilities are zero,
we do not see their difference.
For saying that 
$ A $ and $ \alpha(M) $ are {\it equal in measurements},
it is necessary and sufficient
that their spectral values appearing with nonzero probability coincide.

Ozawa~\cite{Ozawa2006} has formulated the notion of perfect correlation
that characterizes the equality of two observables in measurement.
Suppose that 
two self-adjoint operators $ A $ and $ B $ on a Hilbert space $ \Hilbert $
have spectral decompositions
\begin{equation}
	A = \int \lambda \, E^A ( d \lambda),
	\qquad
	B = \int \lambda \, E^B ( d \lambda)
\end{equation}
with their respective projection measures $ E^A $ and $ E^B $.
It is said
that two observables $ A $ and $ B $ are 
{\it perfectly correlated in a state} $ \psi \in \Hilbert $ 
if the equation
\begin{equation}
	E^A ( \Delta ) \psi =
	E^B ( \Delta ) \psi 
	\label{perfect correlation}
\end{equation}
holds for an arbitrary Borel subset $ \Delta \subset \R $.
Suppose that this relation (\ref{perfect correlation}) holds
and that
the measurements of $ A $ and $ B $ are performed on the state $ \psi $.
When the outcome of $ A $ is in the range $ \Delta $,
the outcome of $ B $ is also in $ \Delta $,
and vice versa.
This property justifies 
calling the relation (\ref{perfect correlation})
the {\it perfect correlation of $ A $ and $ B $ in $ \psi $}.
This relation is denoted as
\begin{equation}
	A \equiv_\psi \! B.
	\label{Ozawa}
\end{equation}
Ozawa proved that the perfect correlations satisfy 
(i) the reflexive law:
$ A \equiv_\psi \! A $,
(ii) the symmetric law:
$ A \equiv_\psi \! B \Rightarrow B \equiv_\psi \! A $,
(iii) the transitive law:
$ A \equiv_\psi \! B, \, B \equiv_\psi \! C \Rightarrow A \equiv_\psi \! C $.
Hence the perfect correlation is an equivalence relation.
In this paper we call it {\it Ozawa equality}.

This equality can be expressed in terms of the GNS construction~\cite{Segal1947}
(GNS is an abbreviation for Gel'fand-Na\u{\i}mark-Segal).
A state $ \omega $ associated to the vector $ \psi \in \Hilbert $
is a linear functional
\begin{equation}
	\omega ( A ) := \bra \psi | A | \psi \ket
\end{equation}
for $ A \in \vect{B} ( \Hilbert ) $,
that is the set of all bounded operators on $ \Hilbert $.
Restricting the state $ \omega $
on the algebra $ \G (A,B) $ generated by $ A $ and $ B $,
and using the GNS procedure,
we can construct a representation $ \pi_\omega $ of the algebra $ \G (A,B) $.
Ozawa himself proved~\cite{Ozawa2006} that the perfect correlation (\ref{perfect correlation}) 
is equivalent to the equality of the GNS-representing operators
\begin{equation}
	\pi_\omega (A) = \pi_\omega (B).
	\label{GNS-Ozawa equality}
\end{equation}

For becoming familiar with the idea of Ozawa equality,
let us examine the following simple example.
Suppose that we have two operators $ A $, $ B $
and a state vector $ \psi $ such as
\begin{equation}
	A = 
	\begin{pmatrix}
	a_{11} & a_{12} & 0 & 0 \\
	a_{21} & a_{22} & 0 & 0 \\
	0 & 0 & a_{33} & a_{34} \\
	0 & 0 & a_{43} & a_{44} 
	\end{pmatrix}
	\!, \quad
	B = 
	\begin{pmatrix}
	b_{11} & b_{12} & 0 & 0 \\
	b_{21} & b_{22} & 0 & 0 \\
	0 & 0 & b_{33} & b_{34} \\
	0 & 0 & b_{43} & b_{44} 
	\end{pmatrix}
	\!, \quad
	\psi = 
	\begin{pmatrix}
	c_{1} \\ c_{2} \\ 0 \\ 0 
	\end{pmatrix}
\end{equation}
with nonzero complex numbers $ c_1, c_2 $.
The GNS procedure associated to the state $ \psi $ yields
\begin{equation}
	\pi_\omega (A) = 
	\begin{pmatrix}
	a_{11} & a_{12} \\
	a_{21} & a_{22} 
	\end{pmatrix}
	\!, \quad
	\pi_\omega (B) = 
	\begin{pmatrix}
	b_{11} & b_{12} \\
	b_{21} & b_{22} 
	\end{pmatrix}.
	\label{GNS}
\end{equation}
So, it can happen that 
$ A \equiv_\psi B $
even when $ A \ne B $.
This result is interpreted as follows.
The probability of emergence of eigenstates associated to 
the right-lower blocks of the matrices $ A $ and $ B $
is zero in the state $ \psi $,
and hence 
these right-lower blocks exhibit no measurable effects.
Thus, when we are concerned with 
the measured values in the state $ \psi $,
it is justified to discard these irrelevant parts
and to leave only the relevant parts as (\ref{GNS}).

Using the Ozawa equality we can characterize the equality
between the quantity to be measured indirectly 
and the quantity to be read out directly.
The initial state vectors of the object and the apparatus
are denoted as $ \psi $ and $ \xi $, respectively.
A precise measurement of the object quantity $ A $
by the meter quantity $ M $ via the measurement process $ \alpha $
from the initial state $ \nu = \psi \otimes \xi $ of the composite system
is characterized by the Ozawa equality
\begin{equation}
	A \equiv_\nu \alpha(M).
	\label{precise measurement}
\end{equation}
Then the observed values of $ A $ and $ \alpha(M) $ always coincide.

\subsection{Main theorem}
We have introduced the three notions:
isolated conservation law,
covariant indicator,
and Ozawa equality.
Combining them we deduce the general superselection rule.

{\it Theorem:
If a measurement process $ \alpha $ admits 
the isolated conservation law associated to the group action of $ G $,
and if an observable $ A $ of the object system 
is precisely measurable by a covariant indicator $ M $ 
in the sense of the Ozawa equality
in an arbitrary object state $ \psi $,
then the quantity $ A $ is $ G $-invariant. 
Namely, the measureable quantity must satisfy}
\begin{equation}
	\sigma_g A = A 
	\label{superselection rule theorem}
\end{equation}
{\it for arbitrary $ g \in G $.}
Let us rephrase the above statement;
a quantity $ A $ that is measurable 
by a $ G $-covariant indicator via a measurement process
preserving the $ G $-symmetry of the object system
must be a $ G $-invariant quantity.
If the group action is generated by $ J $ as 
$ \sigma_s (A) = e^{iJs / \hbar} A \, e^{-iJs / \hbar} $,
then the isolated conservation law $ [ J, H ] = 0 $
implies the superselection rule $ [ J, A ] = 0 $.

{\it Proof:}
Under the assumption of the theorem,
we have a measurement scheme $ ( \alpha, A, M, \nu ) $
and a group action $ ( G, \sigma, \tau ) $
that satisfy 
the Ozawa equality
$ \sigma_g A \equiv_\nu \alpha ( \tau_g M ) $,
the covariance of the indicator
$ \alpha ( \tau_g M ) =  \sigma_g ( \alpha M) $,
and 
the isolated conservation law 
$ \sigma_g \circ \alpha = \alpha \circ \sigma_g $.
Then we have
\begin{equation}
	\sigma_g A
	\equiv_\nu
	\alpha ( \tau_g M) 
	\equiv_\nu
	\sigma_g ( \alpha M )
	\equiv_\nu
	\alpha( \sigma_g M ).
	\label{calculation}
\end{equation}
Note that the Ozawa equality satisfies transitivity.
Since 
$ M = \1 \otimes M $ and
$ \sigma_g \1 = \1 $
for the identity element $ \1 $ of the object algebra,
\begin{eqnarray}
	\alpha( \sigma_g M )
	= \alpha( \sigma_g ( \1 \otimes M ) )
	= \alpha( \1 \otimes M ).
	\label{almost end}
\end{eqnarray}
Therefore, the equalities appearing in (\ref{calculation})
do not depend on $ g \in G $, and hence
\begin{equation}
	\sigma_g A \equiv_\nu A 
	\label{ozawa invariant}
\end{equation}
holds for the state vector $ \nu = \psi \otimes \xi $.
The operator $ A $ is defined in the Hilbert space $ \Hilbert $ of the object system.
We have assumed that
the state vector $ \psi \in \Hilbert $ can be chosen arbitrarily.
(This requirement is not an excessive demand.
In usual experiments,
the initial state $ \xi $ of the apparatus is fixed
but various initial states $ \psi $ of the object are put in.)
Therefore, the equality (\ref{ozawa invariant})
must hold for an arbitrary vector $ \psi \in \Hilbert $.
Thus we reach the conclusion that
the operator $ A $ itself must be $ G $-invariant:
\begin{equation}
	\sigma_g A = A.
	\label{invariant}
\end{equation}
{\it End of proof}.

Here we put a comment.
We proved the theorem under the assumption that 
$ A $ and $ \alpha(M) $ 
are covariant and 
precisely equal in the sense of the Ozawa equality.
We can relax this assumption 
and replace it by the requirement of 
covariance and perfect correlation between 
the probability operator-valued measures (POVMs) 
associated to the object observable and the meter observable.
Under that assumption on POVMs,
we will reach an almost same conclusion as (\ref{invariant}).

\section{Implications}
In the following we will discuss physical implications 
of the superselection rule from the viewpoint of measurement theory.

\subsection{Wigner-Araki-Yanase-Ozawa theorem}
The object system has
a quantity $ J $ generating the symmetry group of the system
and also has 
a quantity $ A $ obeying a nontrivial transformation rule under the group action.
Our theorem tells that
we cannot measure the quantity $ A $
via a measurement process that conserves $ J $.
By contraposition,
disturbance of $ J $ is inevitable in any measurement of $ A $.
This interpretation of the theorem reminds us the uncertainty relation.
Let us examine this point.

The limitation on accuracy of measurements under conservation laws
is known as the Wigner-Araki-Yanase (WAY) theorem~\cite{Wigner1952, Araki1960},
which states that
a precise measurement of a quantity
that does not commute with an additive conserved quantity
is impossible.
Ozawa~\cite{Ozawa2002} reformulated the WAY theorem 
and proved the quantitative relation 
\begin{equation}
	\varepsilon ( A )^2
	\ge
	\frac{ \big| \bra [ A, J_1] \ket \big|^2 }
	{ \: 4 \{ \sigma (J_1)^2 + \sigma (J_2)^2 \} \:},
\end{equation}
where 
$ \varepsilon ( A ) := \sqrt{ \bra ( \alpha(M) - A )^2 \ket } $ 
is the error of measurement of $ A $
and $ \sigma ( J ) := \sqrt{ \bra ( J - \bra J \ket )^2 \ket } $
is the standard deviation of $ J $.
The quantity $ J_1 $ belongs to the object system
while the quantity $ J_2 $ belongs to the apparatus.
It is assumed that
their sum $ J = J_1 + J_2 $ is conserved during the measurement process.
It is also assumed that 
the meter observable $ M $ commutes with $ J_2 $.

In our setting, it is assumed that
the quantity $ J_1 $ alone is a conserved quantity.
The quantity $ J_2 $ can be defined to be identically zero 
and hence we have $ \sigma (J_2) = 0 $.
An initial state such that $ \sigma (J_1) = 0 $ can be prepared.
If $ \bra [ A, J_1] \ket \ne 0 $,
the error $ \varepsilon ( A ) $ diverges
and the measurement fails to make sense.
This consequence resembles the superselection rule.
Thus, the superselection rule can be regarded 
as the strongest version of the Wigner-Araki-Yanase-Ozawa theorem.

However, our derivation of the superselection rule
elucidates the role of covariance 
and clarifies the meaning of measurability.
Suppose that the transformation group of $ A $ is generated by $ J_1 $.
For accomplishing a relevant measurement,
it is desirable that the value of the meter $ \alpha(M) $ varies
when the value of the object quantity $ A $ varies.
This is the requirement of covariance.
But
the covariant correlation cannot be made
via a measurement process that conserves $ J_1 $.
In this sense, $ A $ is non-measurable.
In our view,
the quantity $ A $ is non-measurable
not because the measurement error is infinite
but because 
the value of the meter observable $ \alpha(M) $ cannot follow
the value of the object quantity $ A $.
The conservation law of $ J_1 $ obstructs making the correlation 
between $ A $ and $ \alpha(M) $.

\subsection{Charge superselection rule}
A typical example of superselection rule is the charge superselection rule,
which follows the $ U(1) $ global symmetry.
For bosonic or fermionic creation and annihilation operators 
$ A_j^\dagger $ and $ A_j $, 
the number operator
\begin{equation}
	N := \sum_{j=1}^n A_j^\dagger A_j,
\end{equation}
which is called the charge, 
generates the unitary operator
\begin{equation}
	U_\theta := e^{iN \theta}
	\label{total number}
\end{equation}
for $ \theta \in \R $ and implements the gauge transformation
\begin{equation}
	\sigma_\theta (B) := U_\theta B U_\theta^\dagger,
	\qquad
	\sigma_\theta (A_j) = e^{-i \theta} A_j,
	\qquad
	\sigma_\theta (A_j^\dagger) = e^{i \theta} A_j^\dagger.
	\label{global U(1)}
\end{equation}

Assume that the number operator $ N $ is an isolated conserved quantity.
Namely, assume that
any physically realizable measurement process $ \alpha $ preserves
$ \alpha(N) = N $.
Then the superselection rule tells that
both the self-adjoint component 
$ \frac{1}{2} ( A_j + A_j^\dagger ) $
nor the anti-self-adjoint component 
$ \frac{1}{2i} ( A_j - A_j^\dagger ) $
of the annihilation operator are non-measurable.
However, since the product
$ A_j^\dagger A_k $ is gauge invariant,
\begin{equation}
	\frac{1}{2} ( A_j^\dagger A_k + A_k^\dagger A_j ),
	\qquad
	\frac{1}{2i} ( A_j^\dagger A_k - A_k^\dagger A_j )
\end{equation}
are measurable quantities.
For superconductivity
$ A_j $ represents the Cooper condensate
while for superfluidity
$ A_j $ represents the Bose-Einstein condensate.
Although $ A_j $ itself is not measurable,
a contact of two superconductors defines 
a measurable product $ A_j^\dagger A_k $.
For example, 
the Josephson current is a function of $ A_j^\dagger A_k $,
which can be interpreted as a function 
of the phase difference of complex Cooper condensates.

\subsection{Color superselection rule}
Let us discuss an implication of the superselection rule
for the non-abelian gauge theory, for example, QCD.
We do not attempt 
to provide a fully developed analysis of the color confinement problem here.
We would like to ask the reader 
to permit the presentation of our immature idea.

The color charges are generators of the $ SU(3) $ symmetry,
which is the unbroken rigorous symmetry of the microscopic world.
Thus, the color charges are subject to an isolated conservation law.
Hence, the superselection rule tells that
any measurable quantity must commute with the color charges.
In other words, a measurable quantity must be colorless.
Since $ SU(3) $ is a non-abelian group,
the color charges themselves do not commute with each other.
Therefore, the color charges are non-measurable.
By the same reason, quark and gluon fields are un-observable.
This is a possible explanation of the color confinement.

The color superselection rule can be compared with the charge superselection rule.
The $ U(1) $ symmetry of QED is also unbroken rigorous symmetry.
Thus, the electric charge is an isolated conserved quantity of the microscopic world.
The superselection rule tells that 
charged complex fields are un-observable.
However, since $ U(1) $ is an abelian group,
the electric charge commutes with itself.
Therefore, the electric charge is measurable.

As well known,
quantization of fields subject to local gauge symmetry is highly nontrivial.
The quantum theory of the non-abelian gauge field
involves various subtle ingredients
like ghost fields, auxiliary fields, an indefinite-metric space, the BRS condition, and so on.
Strocchi~\cite{Strocchi1976} showed 
that every observable is a color singlet as a consequence of locality.
Ojima~\cite{Ojima1978} also investigated 
the observability condition of physical quantities using the BRS symmetry.
However, we do not yet have a decisive solution of the confinement problem.
Although our consideration in the present form seems not immediately applicable
to the quantum theory of gauge fields,
we provided a standpoint, at least, 
from which 
the confinement problem is viewed as a subject of measurement theory.

On the other hand,
by the color confinement physicists usually mean
not only that colored quantities are non-detectable
but also that quarks and gluons are confined in hadrons.
Namely, the confinement problem includes also
the problem of bound states of strongly interacting particles.
This aspect is a matter of genuine dynamics
and is out of the scope of measurement theory.

\subsection{Angular momentum}
The color superselection rule can be compared with
the angular momentum conservation law.
Angular momenta are generators of the rotation group $ SO(3) $,
which is a non-abelian group.
However, we can measure angular momenta of various microscopic systems;
experimentalists measure 
angular momenta of atoms or nuclei by using
the Zeeman effect or the Stern-Gerlach experiment setting.
They measure also spin angular momenta of photons
by using polarization filters or birefringent media.
These facts give a rise of a question;
why does not the superselection rule associated to the rotational symmetry
prohibit measurement of angular momenta?

We can measure angular momenta
because the rotational symmetry is spontaneously broken
in the macroscopic world.
For example, a molecule of water has a non-spherical shape.
Shapes of carbohydrate molecules and protein molecules 
are not rotationally invariant, either.
As another example,
magnetization of bulk of iron breaks the rotational symmetry.
In the macroscopic world,
there are a lot of objects that have rotationally asymmetric shapes.
Rotationally asymmetric objects can carry nonzero angular momenta.
Therefore, the conservation of angular momenta
is not closed in the microscopic world.
An interaction can transfer angular momenta between
a microscopic system and a macroscopic system.
Actually, all of the experimental settings
for measuring angular momenta of microscopic systems
break the rotational symmetry by applying
asymmetric external fields on microscopic systems.
The spontaneous breaking of the rotational symmetry
allows the existence of rotationally asymmetric macroscopic objects 
and
enables the measurements of angular momenta by macroscopic apparatus.

On the other hand,
the color $ SU(3) $ is not spontaneously broken.
Therefore, the color superselection rule continues to hide
color charges from measurements.
This argument for justifying the color confinement may seem a tautology;
the color is invisible from the macroscopic world
because the color symmetry has not been broken spontaneously
and there are no macroscopic objects that carry color charges.
But the non-existence of colored objects in the macroscopic world
sounds just a rephrasing of the color confinement.

However, this argument is not a tautology.
Our derivation of superselection rules tells a mechanism of the superselection rules;
the symmetry of measurement process prevents 
the indicator from moving sensitively to a change of the quantity 
that obeys a nontrivial transformation law under the action of the symmetry group.
This also tells an approach for measuring the quantity 
that obeys the nontrivial transformation law;
we can measure the quantity by breaking the symmetry explicitly.
For example, in the Stern-Gerlach experiment setting,
the spin of an atom is measured by applying an
inhomogeneous magnetic field,
which breaks the rotational symmetry explicitly.
In the case of superconductivity,
the phase of a Cooper condensate can be measured
by bringing another Cooper condensate 
and by making a Josephson junction between them.
The second condensate 
exchanges Copper pairs with the first one
and breaks the isolated conservation law 
of the number of Cooper pairs of the first system.
Then the relative phase of the two condensates
can be measured by Josephson current.
Similarly, contact of color superconducting objects 
will enable measurement of their relative color.
In this manner, 
by understanding superselection rules 
from the viewpoint of measurement theory,
we can find a method to extend the class of measurable quantities.

\section{Conclusion}
In this paper,
we reviewed the formulation of superselection rule,
which restricts the class of measurable physical quantities
by requiring them to commute with the superselection charge.
We examined two examples,
in which the momentum conservation law and the position measurement
are incompatible.
In other words, the translational symmetry 
prevents the meter from moving covariantly to the position of the object.
The analysis of these examples told a lesson that
symmetry of measurement process restricts the class of feasible measurements.
For accomplishing a meaningful measurement
it is necessary to make a covariant correlation between
the quantity of an object and the indicator of an apparatus.
We introduced the three basic notions,
isolated conservation law,
covariant indicator,
and Ozawa equality, to prove the theorem;
if a measurement process preserves the symmetry of the object system,
a quantity measurable by a covariant indicator 
should be invariant under the symmetry group action.
This theorem justified superselection rules 
from the viewpoint of measurement theory.
The implication of the charge superselection rule was discussed.
It was also argued that
invisibility of colored quantities can be understood 
as a consequence of the non-abelian color symmetry.
It was noted that
angular momenta are also conserved quantities associated to
the non-abelian $ SO(3) $ symmetry
but they are measurable in experiments.
Spontaneous breaking of the rotational symmetry allows
existence of rotationally asymmetric  macroscopic objects 
that can exchange angular momenta with microscopic objects.
Thus, conservation of angular momenta 
is not closed in the microscopic world.
Hence the superselection rule is not applicable to the rotational symmetry
and we can measure microscopic rotational variables,
in particular, a spin component of a particle.
This consideration suggests a method for measuring a quantity
that obeys a nontrivial transformation law under symmetry operations;
the indicator can move covariantly to the object quantity
if we use a measurement process which breaks the symmetry,
for example, by applying external field 
or by bringing another subsystem 
and allowing an interaction that breaks isolation of the object.
These considerations are not just rephrasing of superselection rules;
they tell a mechanical foundation of superselection rules
and also a method for overcoming superselection rules.

We showed that
the isolated conservation law defines a superselection rule 
and hence defines 
the class of microscopic quantities that are measurable by outside observers.
It may be possible to say that
the isolated conservation law defines 
a border between
the microscopic object world and the macroscopic observer world.
The degree of isolation of the object system is variable.
Thus the class of measurable quantities can vary 
depending on available measurement interactions.
These considerations tell that the superselection rules are not absolute rules.

Finally, we would like to mention a view of the physical world
brought by the study of superselection rules.
Nature has a 
hierarchical structure like 
quarks, hadrons, atoms, molecules, polymers, condensed matters, cells, life, and so on.
It can be said that the hierarchical structure is based on nesting of isolations.
The conservation laws of colors, quark flavors, lepton flavors, chiral symmetry,
isospins, electric charges, angular momenta, and linear momenta
define various levels of isolations.
For example, the colors are isolated and conserved in hadrons, 
the isospins are isolated in nuclei, and so on.
Some of them are rigorous unbroken symmetries,
some are broken,
and the others are approximate symmetries.
The scales of symmetry breakings also spread over 
from the electroweak scale to the molecular scale.
Each level of isolated symmetry or broken symmetry corresponds 
each hierarchy of nature.
In this view, it is recognized that
the division between the micro and the macro is not fixed
but there are various micro-macro strata
which are marked by the superselection rules.
This view seems in harmony with Anderson's view 
that symmetry breakings generate each hierarchy of nature~\cite{Anderson1972}
and Ojima's concept of the Micro-Macro duality
on the bi-directional functions of the micro and the macro physics~\cite{Ojima2005}.

\section*{Acknowledgements}
I would like to express my gratitude to Dr. I. Ojima and the members of his group,
from discussions with whom 
I became conscious of significance of the algebraic approach in quantum theory.
I thank Dr. M. Ozawa for his scrupulous lectures on measurement theory.
I thank also Dr. A. Sugita and Dr. A. Terai for helpful discussions.
I have been writing a series of articles on quantum mechanics
in Japanese magazine, Rikei e-no Suugaku~\cite{Tanimura2011},
and the writing work gave me a motivation to consider the physical meaning
of the superselection rules.
I thank A. Tomita who gave this opportunity to me.
This work is supported 
by Japan Society for the Promotion of Science, 
Grant Nos.~22540281 
and 22540410. 

\baselineskip 4.9mm 


\begin{thebibliography}{99}
\bibitem{WWW1952}
G. C. Wick, A. S. Wightman, and E. P. Wigner, 
``The intrinsic parity of elementary particles,''
Phys. Rev. {\bf 88}, 101 (1952).

\bibitem{Wightman1995}
A. S. Wightman, 
``Superselection rules; old and new,''
IL Nuovo Cim. {\bf 110B}, 751 (1995).

\bibitem{Hegerfeldt1968}
G. C. Hegerfeldt, K. Kraus, and E. P. Wigner,
``Proof of the fermion superselection rule without the assumption of
time reversal invariance,''
J. Math. Phys. {\bf 9}, 2029 (1968).

\bibitem{Strocchi1974}
F. Strocchi and A. S. Wightman,
``Proof of the charge superselection rule in local relativistic quantum field theory,''
J. Math. Phys. {\bf 15}, 2198 (1974).

\bibitem{Ojima2005}
I. Ojima,
``Micro-Macro duality in quantum physics'',
pp.143-161 
in {\it Proc. Intern. Conf. on Stochastic Analysis, Classical and Quantum},
World Scientific (2005);
arXiv: math-ph/0502038.

\bibitem{Zurek1982}
W. H. Zurek,
``Environment-induced superselection rules,''
Phys. Rev. D {\bf 26}, 1862 (1982).

\bibitem{Zurek1991}
W. H. Zurek,
``Decoherence and the transition from quantum to classical,''
Physics Today {\bf 44}, 36 (1991); 
revised and enlarged version, arXiv: quant-ph/0306072.

\bibitem{Zurek2003}
W. H. Zurek,
``Decoherence, einselection, and the quantum origins of the classical,''
Rev. Mod. Phys. {\bf 75}, 715 (2003).

\bibitem{Machida1980}
S. Machida and M. Namiki,
``Theory of measurement in quantum mechanics:
Mechanism of reduction of wave packet. I, II,''
Prog. Theor. Phys. 
{\bf 63}, 1475 (1980); 
1833 (1980).

\bibitem{Araki1980}
H. Araki,
``A remark on Machida-Namiki theory of measurement,''
Prog. Theor. Phys. {\bf 64}, 719 (1980).

\bibitem{DHR1969}
S. Doplicher, R. Haag, and J. E. Roberts,
``Fields, observables and gauge transformations I, II,''
Commun. Math. Phys. 
{\bf 13}, 1, (1969);
{\bf 15}, 173 (1969).

\bibitem{DR1972}
S. Doplicher and J. E. Roberts,
``Fields, statistics and non-abelian gauge groups,''
Commun. Math. Phys. 
{\bf 28}, 331 (1972).

\bibitem{DR1990}
S. Doplicher and J. E. Roberts,
``Why there is a field algebra with a compact gauge group 
describing the superselection structure in particle physics,''
Commun. Math. Phys. 
{\bf 131}, 51 (1990).

\bibitem{DHR1971}
S. Doplicher, R. Haag, and J. E. Roberts,
``Local observables and particle statistics I, II,''
Commun. Math. Phys. 
{\bf 23}, 199 (1971);
{\bf 35}, 49 (1974).

\bibitem{Sewell2002}
G. L. Sewell,
{\it Quantum Mechanics and Its Emergent Macrophysics},
Princeton University (2002), Section 2.6.5.

\bibitem{Ojima2003}
I. Ojima,
``A unified scheme for generalized sectors based on selection criteria: 
order parameters of symmetries and of thermality and physical meanings of adjunctions,''
Open Sys. Info. Dyn. {\bf 10}, 235 (2003).

\bibitem{Hepp1972}
K. Hepp,
``Quantum theory of measurement and macroscopic observables,''
Helv. Phys. Acta {\bf 45}, 237 (1972).

\bibitem{Ozawa2004}
M. Ozawa,
``Uncertainty relations for noise and disturbance
in generalized quantum measurements,''
Ann. Phys. {\bf 311}, 350 (2004).

\bibitem{Ozawa2006}
M. Ozawa,
``Quantum perfect correlations,''
Ann. Phys. {\bf 321}, 744 (2006).

\bibitem{Neumann1955}
J. von Neumann,
{\it Mathematical Foundations of Quantum Mechanics},
translated from the German edition by R. T. Beyer,
Princeton University Press (1955), Section VI.3.

\bibitem{Segal1947b}
I. E. Segal,
``Postulates for general quantum mechanics,''
Ann. Math. {\bf 48}, 930 (1947).


\bibitem{Holevo}
A. S. Holevo,
{\it Probabilistic and Statistical Aspects of Quantum Theory},
North-Holland (1982), 
Chapter IV. Covariant Measurements and Optimality.

\bibitem{Segal1947}
I.~E.~Segal,
``Irreducible representations of operator algebras,''
Bull. Amer. Math. Soc. {\bf 53}, 73 (1947).

\bibitem{Wigner1952}
E. P. Wigner, 
``Die Messung quantenmechanischer Operatoren,''
Z. Phys. {\bf 133}, 101 (1952).

\bibitem{Araki1960}
H. Araki and M. M. Yanase, 
``Measurement of quantum mechanical operators,''
Phys. Rev. {\bf 120}, 622 (1960).

\bibitem{Ozawa2002}
M. Ozawa,
``Conservation laws, uncertainty relations, and quantum limits of measurements,''
Phys. Rev. Lett. {\bf 88}, 050402 (2002).

\bibitem{Strocchi1976}
F. Strocchi,
``Locality, charges and quark confinement,''
Phys. Lett. {\bf 62B}, 60 (1976).

\bibitem{Ojima1978}
I. Ojima,
``Observables and quark confinement in the covariant canonical formalism of Yang-Mills theory,''
Nucl. Phys. B {\bf 143}, 340 (1978).

\bibitem{Anderson1972}
P. W. Anderson,
``More is Different,''
Science {\bf 177}, 393 (1972).

\bibitem{Tanimura2011}
S. Tanimura,
``Quantum theory in the 21st century,'' 
Rikei e-no Suugaku (Japanese magazine published by Gendai-suugaku-sha),
September, October, and November issues (2011).

\end{thebibliography}
\end{document}